\documentclass[%
 reprint,
 amsmath,amssymb,
 aps,
]{revtex4-2}

\usepackage{graphicx}
\usepackage{dcolumn}
\usepackage{bm}
\usepackage[dvipsnames]{xcolor}
\usepackage{hyperref}
\usepackage{amsmath}
\usepackage{braket}
\usepackage{tabularx}

\begin{document}

\preprint{APS/123-QED}

\title{Revisiting Majumdar-Ghosh spin chain model and Max-cut problem using variational quantum algorithms\\}
\author{Britant}
  \email{britant808@gmail.com}
\author{Anirban Pathak}\email{anirban.pathak@jiit.ac.in}

\affiliation{Department of Physics and Materials Science \& Engineering,
Jaypee Institute of Information Technology, A-10, Sector 62, Noida, UP-201309, India}%



\date{\today}

\begin{abstract}
In this work, energy levels of the Majumdar-Ghosh model (MGM) are analyzed up to 15 spins chain in the noisy intermediate-scale quantum framework using noisy simulations. This is a useful model whose exact solution is known for a particular choice of interaction coefficients. We have solved this model for interaction coefficients other than that required for the exactly solvable conditions as this solution can be of help in understanding the quantum phase transitions in complex spin chain models. The solutions are obtained using quantum approximate optimization algorithms (QAOA), and variational quantum eigensolver (VQE). To obtain the solutions, the one-dimensional lattice network is mapped to a Hamiltonian that corresponds to the required interaction coefficients among spins. Then, the ground states energy eigenvalue of this Hamiltonian is found using QAOA and VQE. Further, the validity of the Lieb-Schultz-Mattis theorem in the context of MGM is established by employing variational quantum deflation to find the first excited energy of MGM.  Solution for an unweighted Max-cut graph for 17 nodes is also obtained using  QAOA and VQE to know which one of these two techniques performs better in a combinatorial optimization problem. Since the variational quantum algorithms used here to revisit the Max-cut problem and MGM are hybrid algorithms, they require classical optimization. Consequently, the results obtained using different types of classical optimizers are compared to reveal that the QNSPSA optimizer improves the convergence of QAOA in comparison to the SPSA optimizer. However, VQE with EfficientSU2 ansatz using the SPSA optimizer yields the best results.
\end{abstract}

\maketitle


\section{\label{sec:level1}INTRODUCTION\protect\\}
Quantum computing allows us to harness the inherent advantages of quantum mechanics which led to the development of new algorithms \cite{wei2022quantum,zoufal2019quantum,bravo2023variational,lubasch2020variational} that have the potential to revolutionize the computational techniques. Among these algorithms, variational quantum algorithms (VQA) are emerging as a type of algorithm that can potentially demonstrate quantum supremacy over classical algorithms even in the noisy intermediate scale quantum (NISQ) devices \cite{preskill2018quantum,CAA+21}. VQAs are the quantum analog of the classical machine-learning models, like neural networks. In fact, VQA is essentially a type of hybrid algorithm that requires parametric quantum circuits, where the parameter optimization is done classically \cite{reiner2019finding}. Classical optimization further helps in keeping the circuit depth low because one can simply do more iterations rather than increasing the circuit depth. It is due to this fact that VQAs perform quite well even for the NISQ era in comparison to other algorithms that are made for fault-tolerant quantum computers.\par Since Grover's algorithm \cite{grover1996fast} has been proposed, endeavors have been made for quantum algorithms to surpass the quadratic speed-up and achieve super-quadratic speed-up \cite{berry2022quantifying,babbush2021focus,dalzell2023mind}. For hybrid algorithms to achieve a beyond quadratic speed, improvement in the algorithmic complexity is important to efficiently implement these algorithms for higher dimensional problems on NISQ devices because it will be unable to counter the computational cost of optimization for classical optimization of parameters since training of parameters is an NP-hard problem \cite{bittel2021training}. The quantum approximate optimization algorithm (QAOA) \cite{fg+2014quantum} is a VQA that promises super-quadratic speed-up for solving combinatorial optimization problems (COPs), but at the same time, there is no conclusive theoretical proof till now (see \cite{BBC+24} and references therein). A COP refers to a specific class of problems where the goal is to find the best solution among a finite number of existing solutions. Each solution belongs to a unique combination of the elements or variables of the problems.\par Recently, VQAs have been rigorously studied and various applications and improvements have been discovered that include parameter introduction \cite{yang2017optimizing}, depth optimization \cite{majumdar2021depth}, number factorization \cite{anschuetz2019variational}, deep learning \cite{verdon2018universal}, variational quantum eigensolver (VQE), quantum chemistry \cite{peruzzo2014variational} and many more. VQE is based on the well-known variational method known as Rayleigh-Ritz approximation technique which is used for finding the ground state of a Hamiltonian. Both of these algorithms (QAOA and VQE) are hybrid algorithms that depend on a classical outer loop for finding the optimized parameters. The Hamiltonian is encoded via unitary time evolution. The evolution time is kept very short unlike adiabatic evolution \cite{aspuru2005simulated}, which requires long time evolutions, making it vulnerable to noise and hence less efficient for NISQ devices. The main difference between QAOA and VQE is how we prepare the parameterized ansatz (it is a trial wave function with an initial guess of a set of parameters). Originally, QAOA was introduced to solve the COPs (Max-cut problem) and VQE was introduced to find a molecular Hamiltonian's ground state energy eigenvalues. However, these algorithms can be used to solve different types of problems whether it is a Hamiltonian of any spin or field interaction model or a classical problem mapped to a Hamiltonian. Most of these problems belong to NP-hard and NP-complete classes and their solutions can be encoded into finding the ground state eigenvalues of the Hamiltonian \cite{lucas2014ising}. Some of the classic examples from different domains are portfolio optimization \cite{rebentrost2018quantum}, constrained portfolio optimization \cite{kerenidis2019quantum}, tail assignment problem \cite{vikstaal2020applying} and job shop problem \cite{amaro2022case}, these problems are COPs that can be solved using QAOA and VQE. The QAOA was designed to find the solution in an unstructured solution space for COPs but recently Matwiejew et al. \cite{matwiejew2023quantum} developed an algorithm named quantum multivariable optimization algorithm (QMOA), an extension of QAOA to exploit the structured solution space for the multivariable continuous functions. This gives a speed-up over Grover's algorithm (the upper limit benchmarking for search algorithms) which finds the solution in an unstructured solution space. At the same time, this extension makes the VQAs a more robust and powerful tool to solve more general problems (not limited to COPs).\par VQAs are also emerging as promising candidates for analyzing the energy spectra of various types of Hamiltonian especially quantum many-body systems which include BCS Hamiltonian \cite{sa2023towards}, Ising model \cite{pelofske2023quantum}, and Heisenberg model \cite{liu2019variational}. Majumdar-Ghosh model (MGM) \cite{majumdar1969next1} is one such spin-chain model that is useful in studying material's properties such as magnetization, specific heat capacity. MGM is also a functional model to study quantum magnets \cite{zhou2017quantum}. Other studies also include investigations on entanglement entropy \cite{chhajlany2007entanglement} and quantum phase transition \cite{liu2011quantum}. The original work by Majumdar and Ghosh \cite{majumdar1969next1} includes eigenvalues of up to a chain of 8 spins, and the succeeding work \cite{majumdar1969next2} includes eigenvalues of a chain of 10 spins these results were obtained using Hulthen's construction procedure. Calculating eigenvalues classically has already been a challenging problem and is considered to be NP-hard. This motivates one to utilize the power of VQAs to solve MGM. Motivating by the above facts, in what follows we first solve MGM (i.e., obtained its ground state and first excited states using VQAs) and compared the efficacy of VQE and QAOA and also that of different ansatzes that may be used in VQE and QAOA. The comparison will lead to a set of conclusions but will not ensure the validity of the conclusions for solving systems other than MGM. This observation motivated us to compare the performance of VQE and QAOA in light of one more computationally important problem.

Since QAOA and VQE both are used interchangeably for solving COPs and finding eigenvalues of a Hamiltonian, it is worth testing the efficacy of these algorithms on the Max-cut problem (explained in detail in Sect. \ref{sec:results}) which is a COP. Problems from different domains (e.g., VLSI circuit design \cite{cho1998fast},  network community detection \cite{newman2006modularity} and Ising formulations \cite{lucas2014ising} have already been mapped to the Max-cut problem. The best known classical algorithm to solve Max-cut problem is the semidefinite programming algorithm \cite{goemans1995improved} for Max-cut. Since the Max-cut problem is NP-hard, a polynomial run time classical algorithm is not possible. Thus, we move towards VQAs (VQE and QAOA), in this work, we have chosen the Max-cut problem to benchmark the QAOA and VQE efficiency to know whether the QAOA has any advantage over VQE in solving a COP problem (as QAOA was especially designed for solving COPs).

The rest of the paper is organized as follows. In Sect. \ref{sec:level2}, basic concepts related to QAOA and VQE are described. Subsequently, in Sect. \ref{sec:MGM}, MGM is described, and the possibilities of using different ansatzes to solve MGM using VQE or QAOA are discussed. In Sect. \ref{sec:results}, the results obtained in the present work are described with specific attention to the ground state energy and the first excited energy of MGM obtained using different VQAs. The results obtained for the Max-cut problem via VQE and QAOA are also reported. It is shown that the obtained results fpr MGM satisfy the Lieb-Schultz-Mattis theorem. Further, a comparison of VQE and QAOA is made on the basis of the results obtained for the same problem with the same number of iterations. Finally, the paper is concluded in Sect. \ref{sec:conclusion}.

\section{\label{sec:level2}QAOA and VQE Basics}
\begin{figure}[th]
    \centering
    \includegraphics[width=\linewidth]{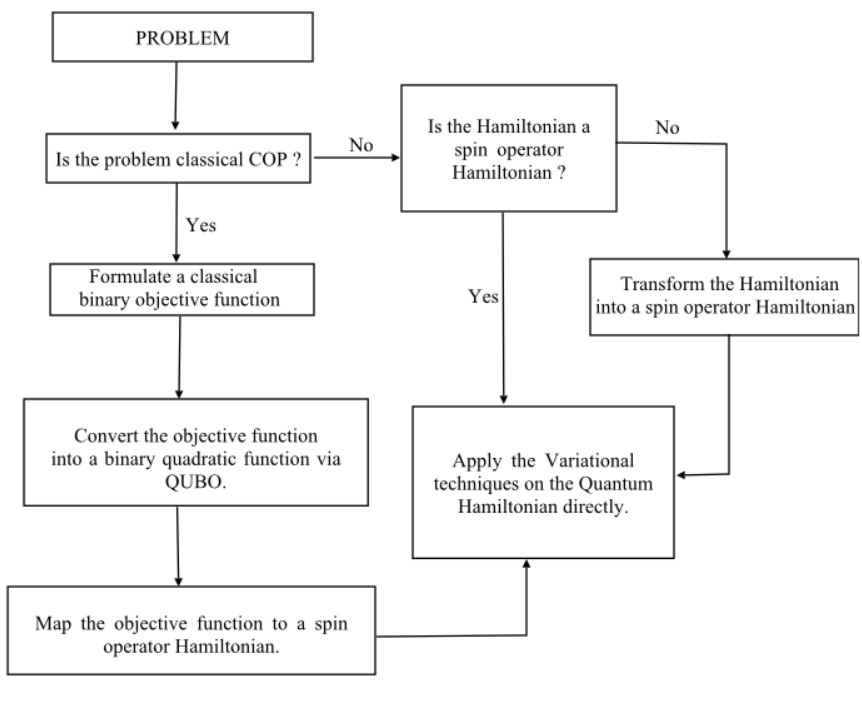}
    \caption{Schematic flow chart of VQA framework.}
    \label{FIG:1}
\end{figure}

As stated above, QAOA was introduced to find the optimal solution for COPs. The Max-cut problem is a typical example of COP. Now, let us see how to obtain the optimal solution for MGM via QAOA. The COPs are encoded into a binary objective or cost function (the function is to be maximized or minimized) that takes binary inputs. Then this function is converted into a quadratic binary function via a method called quadratic unconstrained binary optimization (QUBO) \cite{hammer2012boolean,kochenberger2014unconstrained} (this method can be applied to a large number of NP-complete problems), subsequently this quadratic binary function is mapped to Pauli spin operators to form a spin Hamiltonian. This technique can be used to map any NP-complete problem to QUBO in polynomial time \cite{lucas2014ising}, thus making VQA applicable to different domains. The flow chart illustrating the process is given in FIG. (\ref{FIG:1}).
Once the target Hamiltonian ($H_C$) has been formulated, a parameterized ansatz is prepared as follows:\\
\\
Step 1: Define a unitary operator $U(H_C,\gamma)$, Taking $\gamma$ as one of the parameters.\\
\begin{align}
    U(H_C,\gamma)=e^{-i{\gamma}H_C}
    \label{eq:1}
\end{align}
Step 2: Define an operator $B$ which is a sum of Pauli $\sigma^x$ operators known as a mixing operator.
\begin{align}
    B=\sum_{j=1}^{n}\sigma_j^{x}
    \label{eq:2}
\end{align}

Step 3: Define a unitary operator $U(B,\beta)$, where $\beta$ is one of the parameters.
\begin{align}
    U(B,\beta)=e^{-i{\beta}B}
    =\prod_{j=1}^{n} e^{-i{\beta}\sigma_j^{x}}
    \label{eq:3}
\end{align}
Step 4: Consider that the initial state $\ket{s}$ is a superposition over the computational basis states as given in Eqn.\eqref{eqn:4}, and the parameterized ansatz can be written in the form given in Eqn.\eqref{eqn:5}.
\begin{align}
    \ket{s}&=\frac{1}{\sqrt{2^n}}\sum_{z}^{}\ket{z},\label{eqn:4}\\ 
    \ket{\psi_{\gamma,\beta}}&=U(B,\beta_p)U(H_C,\gamma_p)U(B,\beta_{p-1})U(H_C,\gamma_{p-1})\nonumber\\
    &\quad U(B,\beta_{p-2})U(H_C,\gamma_{p-2})\nonumber\\ 
    &\quad \cdots U(B,\beta_1)U(H_C,\gamma_1)\ket{s}
    \label{eqn:5}
\end{align}
The parameterized ansatz (it acts as a trial wave function) written above is known as $p$-level QAOA ansatz (here p defines the number of repetitions i.e appending the same ansatz with different parameters, each repetition contains a mixer and a unitary operator) which has $2p$ parameters ($\beta_p,\gamma_p,\beta_{p-1},\gamma_{p-1},\cdots,
\beta_1,\gamma_1$). If $p$ is increased, the performance of QAOA gets better, but it enhances the difficulty in implementing it in the NISQ frameworks. As the circuit depth of the ansatz increases on increasing the repetitions,  it makes the circuit noisy due to gate errors \cite{hu2002gate} and decoherence of states \cite{divincenzo1998decoherence}. \\
\\
Step 5: Calculate the expectation value (the cost function) using  Eqn.\eqref{eqn:6} and update the parameters via classical optimizer, and iterate in a loop until it converges to the exact results or the barren plateaus  \cite{mcclean2018barren} (the convergence reaches a saturation state before converging to the solution, this is a very common setback in gradient-based optimizers. However, non-gradient-based optimizers are also not able to solve this problem) occur. 
\begin{align}
    E=\bra{\psi_{\gamma,\beta}}H_C\ket{\psi_{\gamma,\beta}}
    \label{eqn:6}
\end{align}

Unlike QAOA, in VQE we can always choose our ansatz. The most common ansatz for VQE is known as unitary coupled-cluster single double (UCCSD) ansatz and it is prepared as follows.\\
\\
Step 1: The initial state $\ket{s}$ can be chosen according to the problem for faster convergence, it is usually the Hartree-Fock ground state when chemistry Hamiltonian models are concerned, otherwise one can start with a vacuum state. We can write the parameterized ansatz i.e the trial wave function as given in Eqn.\eqref{eqn:7},
\begin{align}
    \ket{\psi_{\theta}}&=e^{T-T^{\dag}}\ket{s}.
    \label{eqn:7}
\end{align}
For a general spin Hamiltonian given in  Eqn.\eqref{eqn:8}, here $h$ is the interaction coefficient between magnetic field and the spins and $J$ is the interaction coefficient among spins.
\begin{align}
    H_C=h\sum_{i\alpha}{}\sigma_{\alpha}^i+J\sum_{ij\alpha\beta}{}\sigma_{\alpha}^i\sigma_{\beta}^{j}, 
    \label{eqn:8}
\end{align}
the cluster operator $T$ is defined as follows
\begin{align}
    T&=T_1+T_2+T_3+\cdots +T_N\\\nonumber
    \text{where}\hspace{0.5cm}
    T_1&=\sum_{i\alpha}{}\theta_i^{r}\sigma_{\alpha}^i\\ \nonumber
    T_2&=\sum_{ij\alpha\beta}{}\theta_{ij}^{\alpha\beta}\sigma_{\alpha}^i\sigma_{\beta}^j.\\\nonumber
\end{align}
Note that the cluster operators need to be transformed (which was given in the original work on VQE \cite{peruzzo2014variational}) into spin operator Hamiltonian via Jordan-Wigner transformation \cite{jordan1993paulische}. However, the above described UCCSD ansatz is not useful for a system containing a large number of spins or qubits as the circuit depth grows rapidly with the size of the system.\\
\\
Step 2: The classical optimization is the same as in QAOA, here the parameter $\theta$ is updated after each iteration till it converges.
\section{Majumdar-Ghosh model and its ground state\label{sec:MGM}}
\begin{figure}[th]
    \centering
    \includegraphics[width=\linewidth]{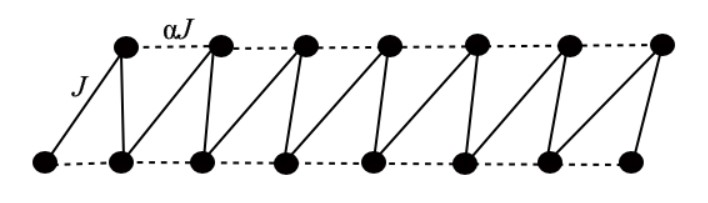}
    \caption{A spin ladder representation of MGM with $J$ and $\alpha J$ } being the interaction strength between nearest neighbor and next neighboring spin, respectively.
    \label{FIG.2}
\end{figure}

\begin{figure}[th]
    \centering
    \includegraphics[width=\linewidth]{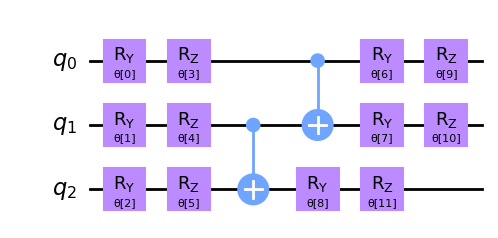}
    \caption{(Color online) Efficient SU2 ansatz (repetition = 1, also \lq reps\rq in qiskit to denote the repetitions of the ansatz) for 3 spins. Here, $R_x, R_y, R_z$ are the single qubit rotation gates that rotate the state of the qubit with $\theta$ degrees about $x,y,z$ axis respectively.}
    \label{Fig.3}
\end{figure}

The MGM is a one-dimensional Heisenberg spin chain model in which we consider interactions not only with the nearest neighboring (NN) spin but also with the next nearest neighboring (NNN) spin, illustrated in FIG. (\ref{FIG.2}). It is one of the most well-known many-body systems whose exact solution is known for the so-called MG-point where $\alpha=\frac{1}{2}$ whose Hamiltonian is given by
\begin{align}
    H=\frac{1}{2}J\sum_{i=1}^{n}\sigma_i\sigma_{i+1}+\frac{1}{2}\alpha J\sum_{i}^{n}\sigma_{i}\sigma_{i+2}
    \label{eq:10}
\end{align}
where $\sigma_i=(\sigma^{x},\sigma^{y},\sigma^{z}$),and the periodic boundary conditions are $n+1\equiv 1$ and $n+2\equiv 2$ (here $n$ is the number of spins in a chain). In this model, the interaction types can be of antiferromagnetic and ferromagnetic nature. If $J>0$ and $\alpha<0$: the NN and NNN interactions will be antiferromagnetic and ferromagnetic, respectively. For simplicity from this point onwards without the loss of generality, we fix $\alpha =-0.1$ and $J=1$. \\
\\MGM is also a useful example of the Lieb-Schultz-Mattis (LSM) theorem \cite{lieb1961two} (a no-go theorem) which states that for an infinite, one-dimensional periodic spin chain, a half-odd-integer spin system has an energy gap (between ground and excited states) of constant$/L$, where $L$ is the length of the chain. As the length grows, the energy gap decreases and eventually, it vanishes as the length of the chain tends to infinity. To calculate the ground state energies we use VQE for large systems and Variational quantum deflation (VQD) \cite{higgott2019variational} for calculating the ground and first excited state for smaller systems only, as the classical optimization becomes very costly if VQD is used for larger systems. The classical optimization complexity also grows as the number of parameters grows in the ansatz. Our choice of ansatz is a hardware efficient ansatz (EfficientSU2- which is IBM's built-in ansatz) \cite{kandala2017hardware} for VQE that takes physical constraints (such as how the qubits are connected) into account. Thus, it takes care of the real hardware while creating the ansatz which leads to a low depth ansatz with a considerably good convergence towards the solution. The term SU2 stands for a special unitary group of the degree of 2, basically, $2\times2$ matrices with determinant 1 such as Pauli matrices (Pauli rotation gates in the ansatz). The EfficientSU2 ansatz for 3 spin chain is given in FIG (\ref{Fig.3}).
\par
\begin{figure}[h]
    \centering
    \includegraphics[scale=0.4]{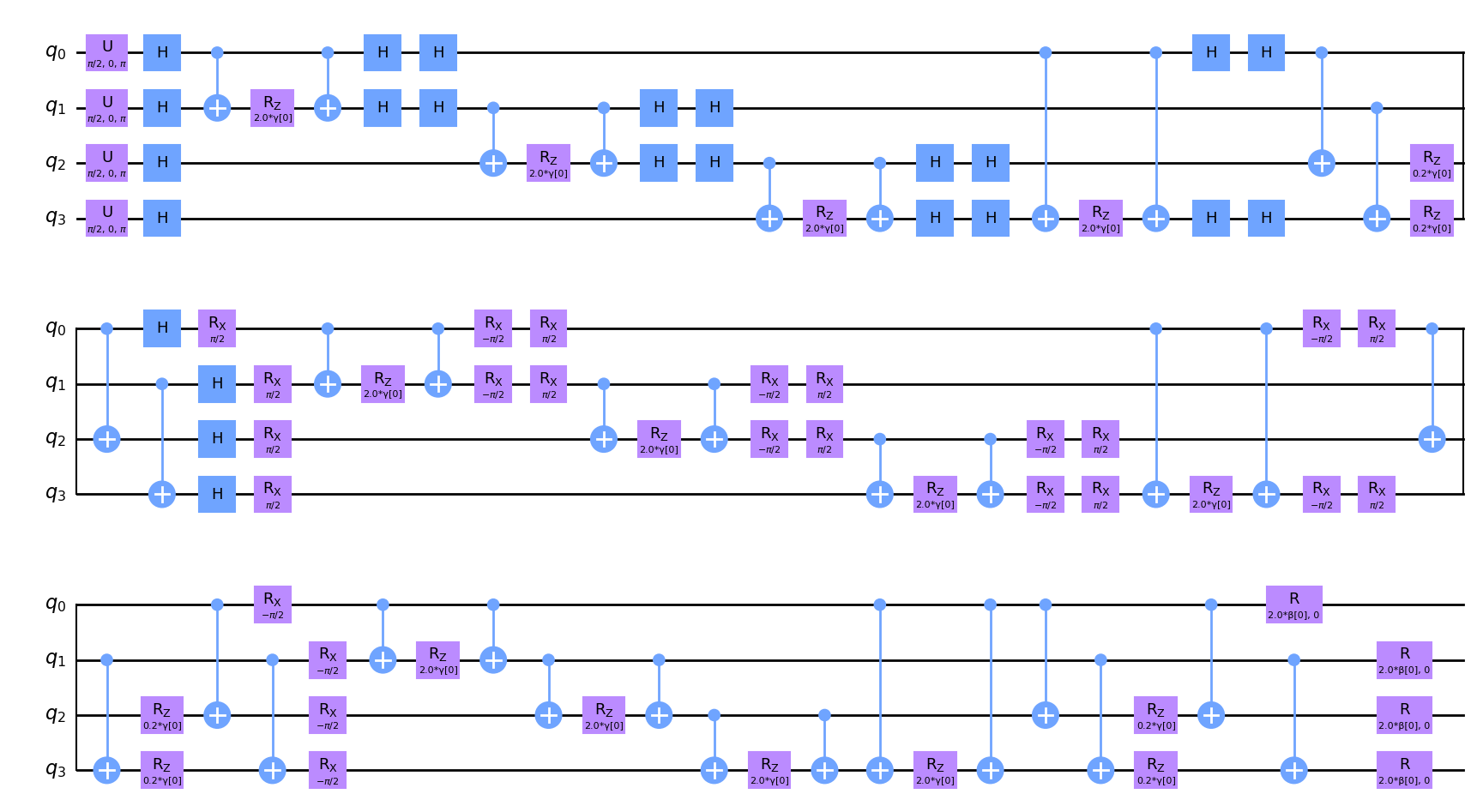}
    \caption{(Color online) QAOA ansatz (reps = 1) for 3 spins. The quantum circuit shown above is a single circuit where the output of the top (middle) panel is to be viewed as the input of the middle (lower) panel. Here, U is a single-qubit rotation gate with 3 Euler angles named as $\theta,\phi,\lambda$.}
    \label{FIG.4}
\end{figure}

 Now, let us look at the standard QAOA ansatz for 3 spins in FIG. (\ref{FIG.4}), which unlike the VQE ansatz takes the Hamiltonian into account while creating the ansatz.
We chose to create these ansatzes for only 3 spins in both cases so that we can show the QAOA ansatz here as it grows rapidly with the number of spins otherwise it would have been difficult to show it. 

We can compare the depth and number of parameters with repetitions of EfficientSU2 ansatz (see Table (\ref{tab.1}) ) and QAOA ansatz (see Table (\ref{tab.2}) ) for 15 spins. If we increase the repetitions, there is a constant increment in the parameters and depth for fixed-length spin. For example, if we take the 15 spins, the EfficientSU2 ansatz for VQE has a depth increment of 4 and a parameters increment of 30 upon increasing 1 repetition. Similarly, QAOA has a depth increment of 391 and parameter increment of 2. QAOA has the disadvantage of creating an ansatz of large circuit depth with less number of parameters thus one has to create the ansatz with more repetitions which leads to an even larger depth.
\onecolumngrid\
\twocolumngrid\
\par To deal with barren plateaus, the ansatz is created with more repetitions (i.e., reps = integer) for bigger systems to get more parameters to train the ansatz circuit so that the gradient of the cost function with respect to parameters does not vanish before the convergence to the desired state occurs.

Eventually, QAOA ansatz is prone to errors for NISQ devices. While VQE with Efficient ansatz is a better option as compared with QAOA, as it creates a lower-depth circuit with more parameters, at the same time it is prone to barren plateaus. More parameters in VQE and QAOA certainly help to get better convergence but it adds to the optimization overhead.\par Our choice of the optimizer is simultaneous perturbation stochastic approximation (SPSA) \cite{spall1998overview} which is useful in the case of noisy simulation. Similarly, sequential least squares programming optimizer (SLSQP)\cite{kraft1988software} is one of the useful optimizers for the non-noisy state vector simulation (IBM's simulator\textunderscore statevector). Keeping the above in mind, in what follows, we will provide an example of the Lieb-Schultz-Mattis theorem and find the ground state energy for MGM using VQE and QAOA. We will also use VQE and QAOA to solve Max-cut problem. While performing these investigations, we will be particularly focused on comparing the performance of VQE and QAOA, and also the choices of ansatz.

\begin{table}
\begin{tabular}{|c|c|c|}
\hline 
repetitions & circuit depth & number of parameters\tabularnewline
\hline 
1 & 18 & 60\tabularnewline
\hline 
2 & 22 & 90\tabularnewline
\hline 
3 & 26 & 120\tabularnewline
\hline 
4 & 30 & 150\tabularnewline
\hline 
5 & 34 & 180\tabularnewline
\hline 
\end{tabular}

\caption{Variation of circuit depth and number of parameters with the number repetitions when EfficientSU2 ansatz is used for 15 spins\label{tab.1}}
\end{table}

\begin{table}
\begin{tabular}{|c|c|c|}
\hline
repetitions & circuit depth & number of parameters\tabularnewline
\hline 
1 & 392 & 2\tabularnewline
\hline 
2 & 783 & 4\tabularnewline
\hline 
3 & 1174 & 6\tabularnewline
\hline 
4 & 1565 & 8\tabularnewline
\hline 
5 & 1956 & 10\tabularnewline
\hline 
\end{tabular}

\caption{QAOA\label{tab.2}Variation of circuit depth and number of parameters with the number repetitions when QAOA ansatz is used for 15 spins. Comparing with Table \ref{tab.1}, we can see that the use of this approach reduces the number of parameters at the cost of a considerable increase in the circuit depth.}

\end{table}


\section{Results\label{sec:results}}
\subsubsection{Ground state energy of Majumdar-Ghosh model using VQE and classical method}
\begin{figure}[h]
    \centering
    \includegraphics[width=\linewidth]{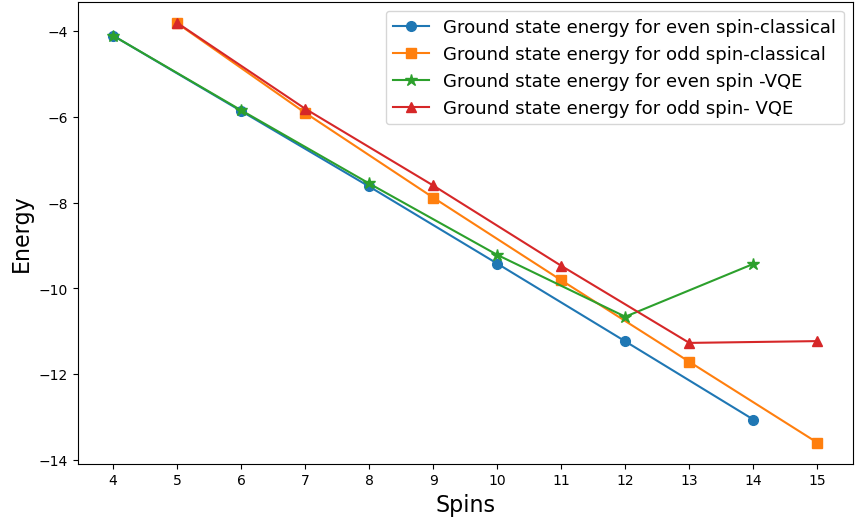}
    \caption{(Color online) Ground state energies comparison from 4 to 15 spins chain. }
    \label{FIG.7}
\end{figure}

In MGM, the smallest chain possible for the NNN interaction to get fully developed is 4 spins, we start with 4 spins and go up to 15 spins. We will show and compare the results obtained from VQE and the results computed classically (using NumPyEigensolver-a qiskit's built-in-function, this function follows an algorithm that transforms the eigenvalue problem to a task to find the roots of a rational function via an iterative numerical approach) in FIG. (\ref{FIG.7}).

The VQE algorithm was performed on simulator\textunderscore statevector using SLSQP optimizer. We kept increasing the depth of EfficientSU2 ansatz and the number of iterations as the number of spins got bigger. For better clarity, we have plotted the even and odd number of spin cases separately since their non-constant slope value is in different ranges. For smaller spins, the values obtained from VQE were quite close to the classically obtained results, and upon reaching 14 and 15 spins the energies started deviating from the classical results (due to the barren plateaus, the results did not converge to the desired value ) which is evident from FIG. (\ref{FIG.7}).\par
\begin{figure}[th]
    \centering
    \includegraphics[width=\linewidth]{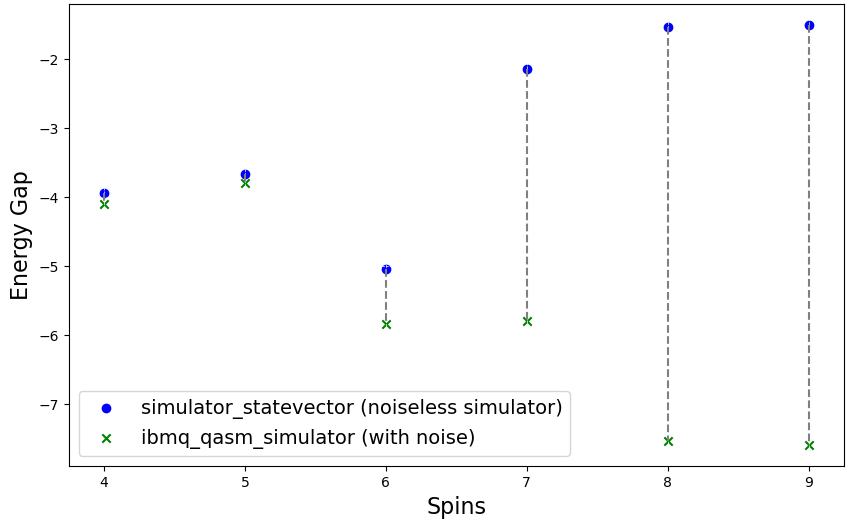}
    \caption{(Color online) VQE results obtained from simulator\_statevector and ibmq\_qasm\_simulator (with noise)  from 4 to 9 spins chain. }
    \label{FIG.8}
\end{figure}

Now, let us compare the results in FIG. (\ref{FIG.8}) obtained from VQE in non-noisy and noisy simulations \cite{wood2020special}.
Again, as we moved towards the higher number of spins the gap between noisy simulation results and non-noisy simulations gets huge. Since we were not opting for any noise-mitigating techniques, we had to stop at 9 spins as the results deteriorated due to noise.

\subsubsection{QAOA vs VQE}
\begin{figure}[h]
    \centering
    \includegraphics[width=\linewidth]{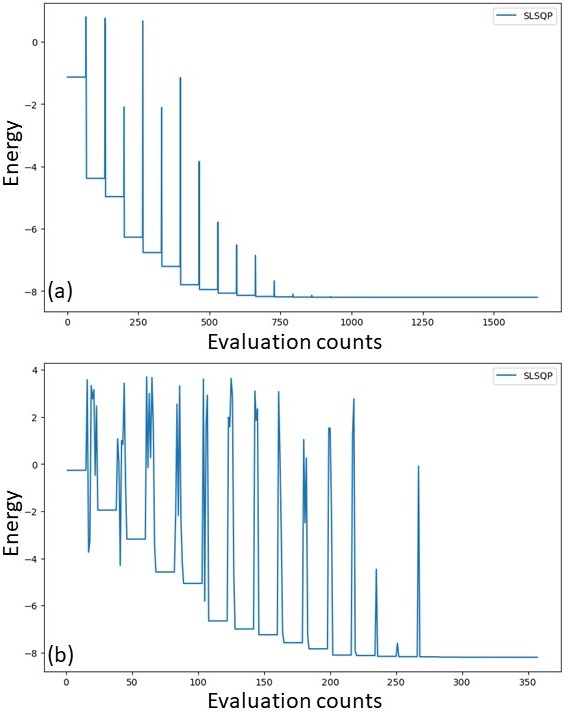}
    \caption{(Color online) (a) VQE (b) QAOA convergence plots for 4 spins in MGM where evaluation counts are the number of times the cost function Eqn.(\ref{eqn:6}) is evaluated in each iteration.}
    \label{FIG.9}
\end{figure}

The benchmarking of VQE and QAOA is done for MGM starting with 4 spins on IBM's non-noisy named simulator\textunderscore statevector. The classical solution for this chain is found to be -4.1 whereas VQE (with EfficientSU2 ansatz with 5 repetitions containing 48 parameters) and QAOA (with 5 repetitions containing 10 parameters) produced -4.099999380749125 and -4.099999884658992 respectively (see FIG. (\ref{FIG.9}) ). All the energy convergence plots in this work have a factor of 2 i.e the energy is shown as twice the actual energy because we did all the simulations by taking out $1/2$ factor out of the Hamiltonian, consider the energy by dividing it by 2. Actual energy is the same as plotted for rest of the plots. Since the results are close to each other, this does not give any certainty of better performance of either algorithm. 
\par
\begin{figure}[h]
    \centering
    \includegraphics[width=\linewidth]{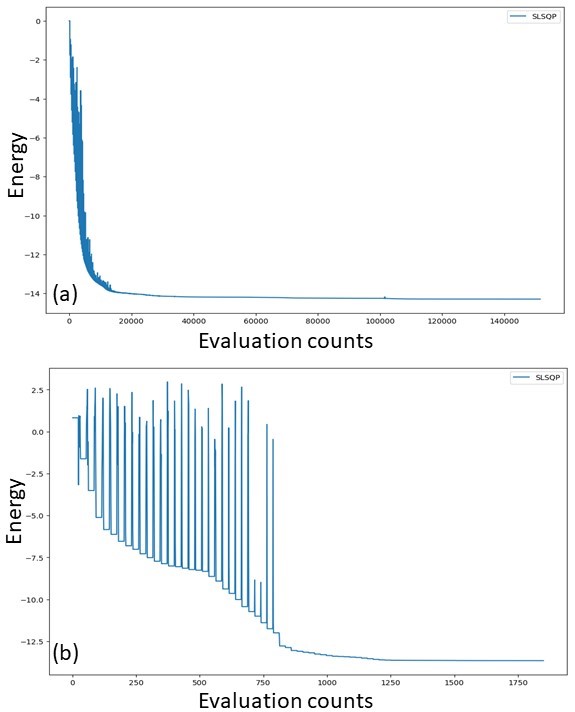}
    \caption{(Color online) (a) VQE and (b) QAOA ($p = 10$, and 20 parameters) convergence plots for 8 spins.}
    \label{FIG.10}
\end{figure}

To get better clarity we increased the chain size to 8 spins. The classical solution, in this case, is -7.62051, the VQE with EfficientSU2 ansatz has 192 parameters and a circuit depth of 51 yielded -7.14720 whereas QAOA with 22 parameters (in QAOA ansatz the number of parameters is always 2 times the repetitions of the ansatz) and circuit depth 540 yielded -6.8247948, the convergence plot is given in FIG. (\ref{FIG.10}). In both the cases, the number of iterations was 1300. It is evident that VQE with lower circuit depth produces better results. However, it comes at the cost of classical optimization as the number of parameters grows significantly with the EfficientSU2 ansatz.
Since the number of parameters in the QAOA ansatz is quite low in comparison with the EfficientSU2 ansatz, it is worth verifying the performance of the QAOA ansatz with more parameters i.e., for higher $p$ or more repetitions. 
\begin{figure}[h]
    \centering
    \includegraphics[width=\linewidth]{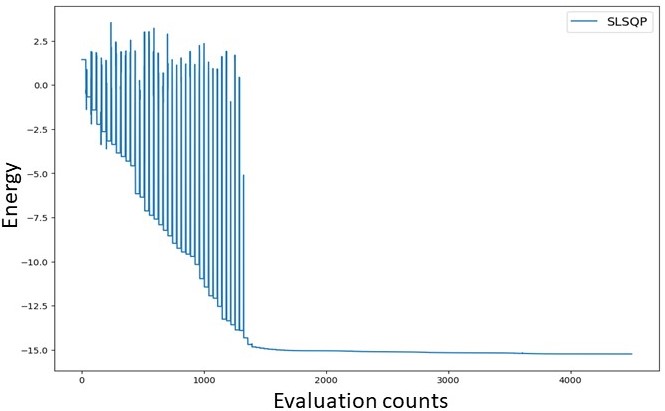}
    \caption{(Color online) QAOA convergence plot for 8 spins with $p = 16$ (32 parameters). }
    \label{FIG.11}
\end{figure}
When the QAOA ansatz with more parameters (32 parameters)  was used for the same number of iterations (1300), it yielded -7.61678 which is quite close to the classical results (see FIG.(\ref{FIG.11})). So QAOA can perform better than VQE even if it contains less number of parameters than VQE. But the circuit depth for QAOA is 2300 which is quite big and impractical for NISQ frameworks (in our case noisy simulations).\par
\begin{figure}[th]
    \centering
    \includegraphics[width=\linewidth]{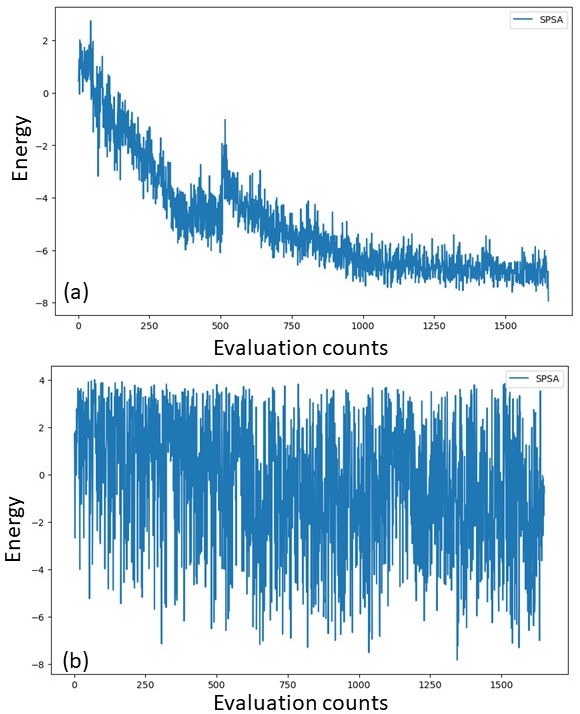}
    \caption{(Color online) (a) VQE (7 reps) and (b) QAOA (5 reps) energy convergence plot for 4 spins in the noisy simulator for 800 iterations.}
    \label{FIG.12}
\end{figure}

Let us analyze the results obtained from VQE and QAOA in noisy simulation (we have used IBM's simulator named ibmq\textunderscore qasm\textunderscore simulator) in FIG. (\ref{FIG.12}). We take the chain of 4 spins and show the energy convergence after each iteration of the classical outer loop using the SPSA optimizer. The classically obtained ground state energy is found to be $-4.10000$, after 800 iterations for both VQE and QAOA, the energies obtained are $-3.96699$ and $-0.37625$, respectively. The Efficient ansatz for VQE was created with 7 repetitions and the QAOA ansatz with 5 repetitions. 
We can observe that QAOA does not guarantee convergence in noisy simulation whereas VQE results were considerably close to the classical result. The circuit depth for QAOA ansatz with 5 repetitions for 4 spins is 335. Eventually, this circuit becomes prone to errors due to the noise.\par
\begin{figure}[th]
    \centering
    \includegraphics[width=\linewidth]{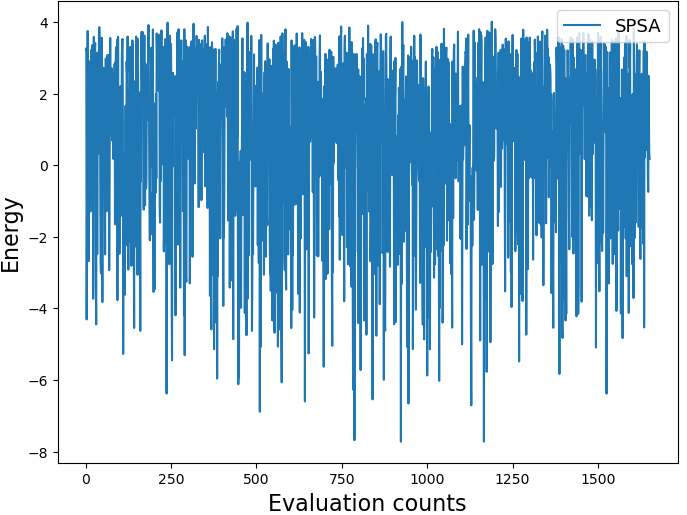}
    \caption{(Color online) QAOA convergence plot with ansatz having 10 repetitions and 800 iterations for 4 spins again in the noisy simulator.}
    \label{FIG.13}
\end{figure}

To verify whether this is happening due to noise or the low number of parameters, we apply QAOA with ansatz having 10 repetitions which is more than before (5 repetitions) and hence involved more parameters. However, it failed to show convergence as shown in FIG. (\ref{FIG.13}), proving that the noise is deteriorating the result and it is not due to the low number of parameters.  

\begin{figure}[th]
    \centering
    \includegraphics[width=\linewidth]{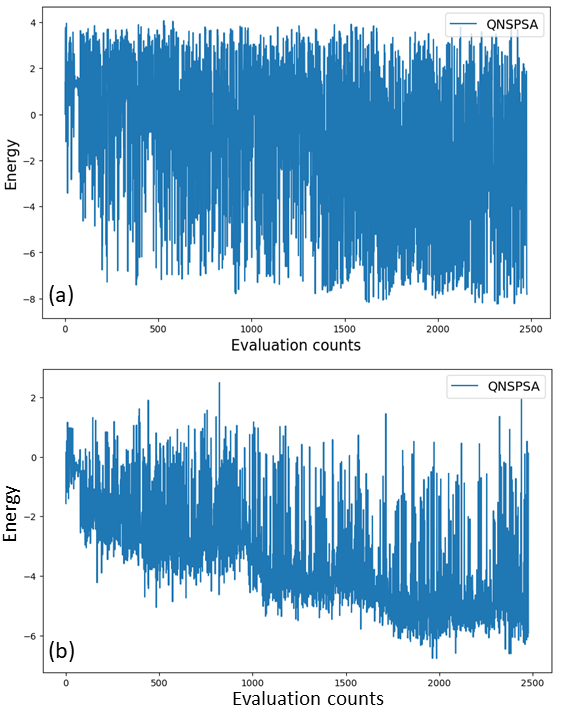}
    \caption{(Color online) (a) QAOA and (b) VQE convergence plot with QNSPSA optimizer for 4 spins.}
    \label{FIG.14}
\end{figure}

Now we use VQE and QAOA (again for 4 spins in noisy simulations) with quantum natural-SPSA (QNSPSA) optimizer which has been reported to perform better than SPSA optimizer in the original paper \cite{gacon2021simultaneous}. We have used QAOA for 800 iterations (with 5 repetitions) which yielded -3.88312 (the classical result is -4.1), it showed better convergence than SPSA for QAOA which is also evident from the convergence plot given in FIG. (\ref{FIG.14}a), QAOA with QNSPSA performed almost equally good as VQE with SPSA. Now we take the EfficientSU2 ansatz for VQE (with 7 repetitions) and perform 800 iterations which yielded -2.96187. So, this optimizer failed to give better results than SPSA for VQE, the convergence plot is given in (\ref{FIG.14}b). Even though QAOA ansatz contains less parameters than VQE ansatz (EfficientSU2) it showed better convergence VQE in noisy simulations with QNSPSA. This suggests that the performance of an optimizer is dependent upon the chosen ansatz and the problem. 
\subsubsection{The Max-cut problem}
\par
\begin{figure}
    \centering
    \includegraphics[width=\linewidth]{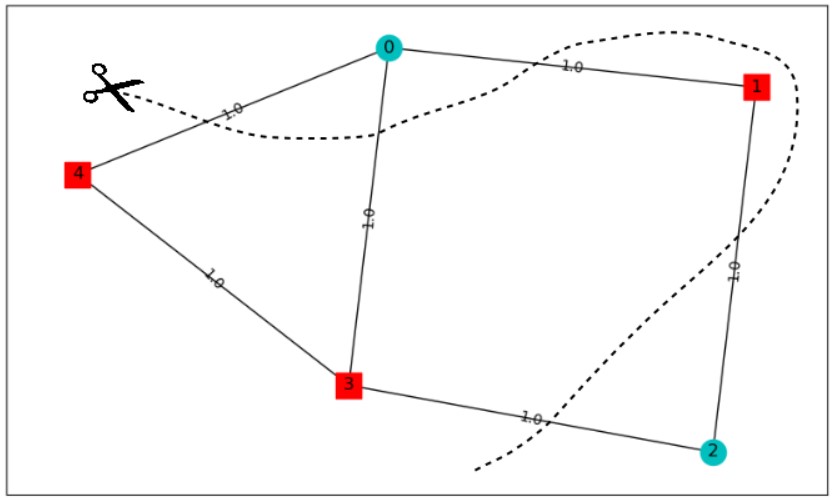}
    \caption{(Color online) A solved Max-cut graph consisting of 5 nodes, where the blue circles and red squares show two parts of the graph after it has been cut.}
    \label{FIG.15}
\end{figure}

 Now let us extend the comparison of QAOA and VQE (and of course, the comparison of the choice of optimizers) to Max-cut problems which is extremely important in its own right. However, the specific purpose of extending the investigation to the Max-cut problem is to check whether the inferences drawn till now are valid only for MGM or are more general in nature.  Specifically, in what follows, we will compare the results of QAOA and VQE on the non-noisy simulator for a Max-cut graph of 17 nodes. A Max-cut problem goal is to find the cut that divides an undirected weighted graph consisting of $n$ nodes into two parts such that the sum of the weight of edges that have been cut is maximum. In FIG. (\ref{FIG.15}), a solved Max-cut has been shown in which the weight of the cut is 5. The blue circles and red squares show two parts of the graph after it has been cut. 

\begin{figure}[h]
    \centering
    \includegraphics[width=\linewidth]{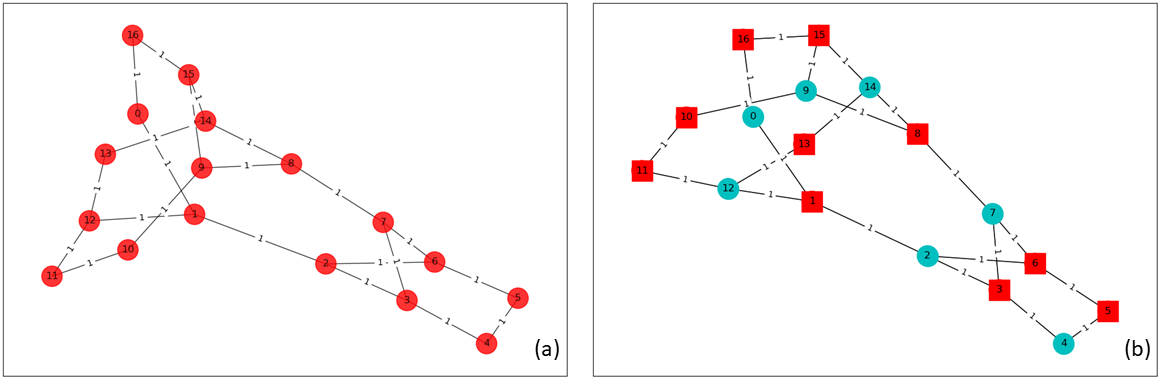}
    \caption{ (Color online) (a) An unweighted unsolved graph consisting of 17 nodes. (b) The same Max-cut graph was solved via the classical brute force approach with the solution objective being 19. The blue circle and red squares show two parts of the graph after it has been cut.}
    \label{FIG.16}
\end{figure}
\begin{figure}[h]
    \centering
    \includegraphics[width=\linewidth]{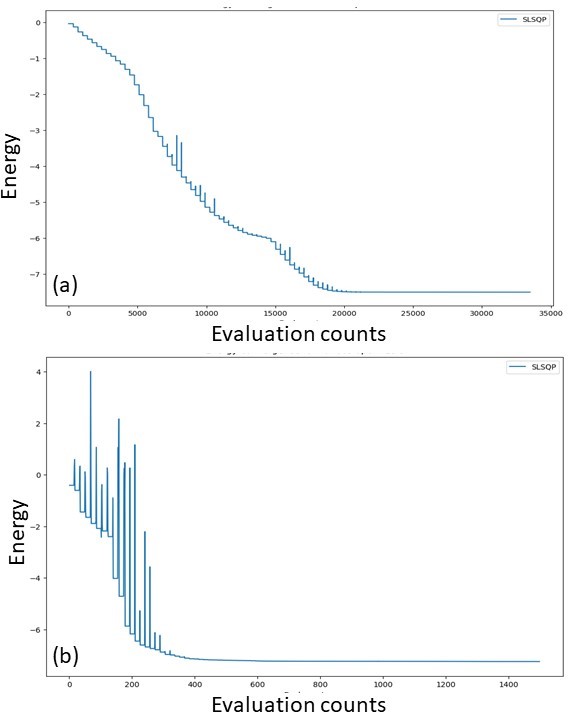}
    \caption{(Color online) (a) VQE (b) and QAOA convergence plot for the Max-cut problem consists of 17 nodes in FIG. (\ref{FIG.16}a) .}
    \label{FIG.17}
\end{figure}

A graph containing 17 nodes (unsolved) is given in FIG. (\ref{FIG.16}a), the corresponding solved Max-cut graph by the classical brute-force approach is given in  FIG. (\ref{FIG.16}b), convergence plot for VQE and QAOA is given in FIG. (\ref{FIG.17}), VQE (with EfficientSU2 ansatz with 9 repetitions and 100 parameters) again showed better convergence for the combinatorial optimization problem having a circuit depth of only 52, giving -18.49986 whereas QAOA with 7 repetitions yielded -18.2472 with a circuit depth of 274.
A similar observation has been made by Amaro et al. \cite{amaro2022case} for the job shop scheduling problem (a combinatorial optimization problem), where VQE performed better than QAOA. This serves as another evidence that QAOA is not very effective for combinatorial optimization problems, for which it was originally intended. Here, VQE wins over QAOA even after having a lower circuit depth again the reason being that QAOA ansatz creates a high-depth circuit with a relatively low number of parameters as compared to EfficientSU2 ansatz. As observed in the case of MGM, an ansatz with a higher number of parameters gives better convergence towards the solution.

\subsubsection{Lieb-Schultz-Mattis theorem for MGM }
\begin{figure}[h]
    \centering
    \includegraphics[width=\linewidth]{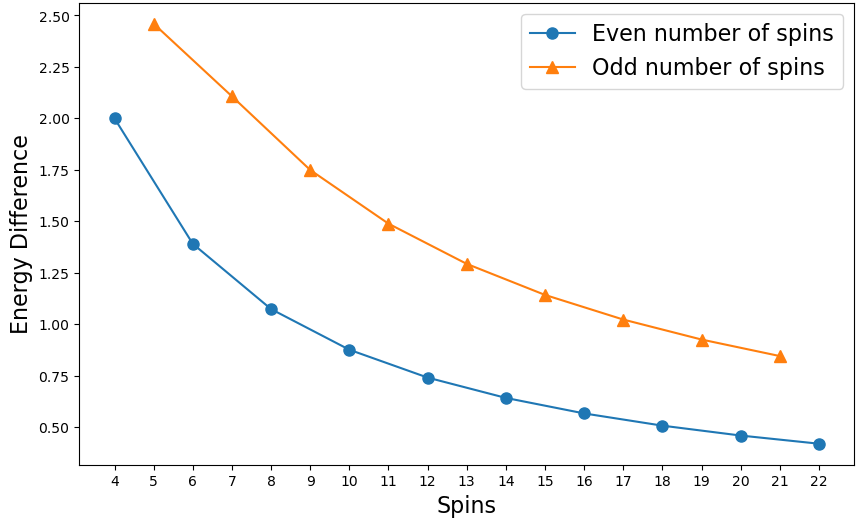}
    \caption{(Color online) The blue line represents the energy difference between the first excited and ground state for the even number of spins. Similarly, the orange line represents the energy difference for the odd number of spins.}
    \label{FIG.18}
\end{figure}

As stated earlier, the energy difference between the first excited state and the ground state starts decreasing and it vanishes as the number of spins in a chain tends to infinity. We may compare the results obtained using NumPyEigensolver and variational quantum deflation (VQD) which is used as a non-noisy simulation. Here, we may note that VQD is an extension of VQE which only gives the ground state of a system, whereas VQD gives excited state energies as well. VQD has been used to solve BCS-Hamiltoinan \cite{sa2023towards} and analyze its energy spectra.  In FIG. (\ref{FIG.18}), we have plotted the classically obtained results for the odd and even number of spins separately since their non-constant slopes are in different ranges, as previously observed while finding the ground state energies. As the number of spins grows in the chain, the energy difference gets smaller which is evident from FIG. (\ref{FIG.18}). 
\par
\begin{figure}[h]
    \centering
    \includegraphics[width=\linewidth]{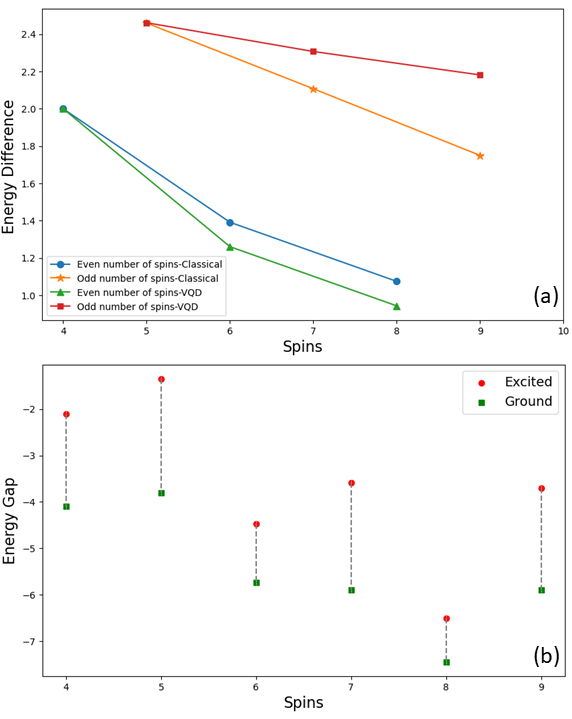}
    \caption{(Color online) (a) Classical vs VQD results up to 9 spins, showing the energy difference between first excited and ground state (b) same VQD results plotted to show the gap (the length of the dotted line represents the gap or the difference) between first excited and ground state energy. }
    \label{FIG.19}
\end{figure}

 Now, let's compare the classical results with the VQD results. In FIG. (\ref{FIG.19} (a), we have plotted the results, and as expected for a lower number of spins the classical and VQD results for even as well as odd number of spins both were close to each other. For VQD, we chose our ansatz to be again EfficientSU2 ansatz and kept increasing the repetitions of the ansatz and the iterations as the number of spins increased in the chain. From the MGM case, we already know that in non-noisy conditions SLSQP works the best, so we chose our optimizer to be SLSQP. As the number of spins in the chain increased, VQD results started deviating from the classical results.
VQD is a useful algorithm to find the energy of excited states, but it converges slowly as the number of parameters required is large, which requires more evaluation per iteration. Eventually, it increases the cost of classical optimization as well. 

\section{Conclusion\label{sec:conclusion}}
In this work, we have studied the application of the QAOA and VQE to solve the MGM and Max-cut problem. Our investigation is focused on understanding the impact of varying circuit depths, number of parameters, and different choices of optimizers and how these factors affect the efficiency of the quantum algorithms. We observed that increasing the circuit depth by repeating the ansatz (which ultimately leads to a circuit with more parameters) in both QAOA and VQE led to improved accuracy (on noiseless simulator\_statevector) in approximating the ground state energy of the MGM as well as optimizing the objective function in the combinatorial optimization problem (specifically in Max-cut problem).
 However, there is a trade-off between errors due to noise and accuracy (in noisy simulations) which requires more parameters hence larger depth. As the circuit depth grows, the required number of quantum gates also increases, affecting the overall accuracy of the solution.
\par We have observed that QAOA can give better convergence in comparison with VQE if the QAOA ansatz contains a comparable or equal number of parameters as in the EfficientSU2 ansatz, but in noisy simulation QAOA always has a low convergence rate in comparison with VQE due to its large circuit depth.
 We have also used the QNSPSA optimizer as Gacon et al. \cite{gacon2021simultaneous} showed that it performs better than the SPSA optimizer in the case of Pauli two-design ansatz (it is prepared with alternating rotation and entanglement layers having an initial layer of $RY(\pi/4)$ gates). Interestingly, in our case, the QNSPSA optimizer is not found to perform better than the SPSA optimizer for VQE, but it is found to improve the convergence in the case of QAOA. We have observed that the performance of an optimizer is dependent on the chosen ansatz. So whether it is a spin Hamiltonian or a combinatorial optimization problem, the optimal choice among optimizers is subjective to the specific problem and ansatz preparation. Whereas among variational methods VQE seems to be better than QAOA in every scenario. Since VQE has the flexibility of choosing ansatz, it can be performed with a lower depth ansatz circuit (like EfficientSU2 ansatz) which provides faster convergence in time and has the advantage of being less prone to error due to noise.
\par As quantum hardware continues to improve, these algorithms hold promise for tackling more complex condensed matter physics problems as well as real-life applications that are NP-hard.
 Further exploration and refinement of these quantum approaches will pave the way for solving challenging optimization problems in general and materials science problems in particular in the near future.
 
\begin{acknowledgments}
Authors acknowledge the support from the Interdisciplinary Cyber Physical Systems (ICPS) programme of the Department of Science and Technology (DST), India, Grant No.: DST/ICPS/QuST/Theme-1/2019/6 (Q46). They also thank Kuldeep Gangwar for some useful discussion.
\end{acknowledgments}

\bibliography{apssamp}

\end{document}